\documentclass{appolb}

\usepackage{graphicx}
\usepackage{amsmath}

\usepackage[T1]{fontenc}
\usepackage[utf8]{inputenc}
\newcommand{\mincas}{{\sf MINCAS}}
\newcommand{\katie}{{\sf KATIE}}
\usepackage{graphicx}
\usepackage{dcolumn}
\usepackage{bm}
\usepackage{epstopdf}
\usepackage{mathrsfs}
\usepackage{amsmath,amssymb,amsfonts,latexsym}
\usepackage{amsmath,bbold}
\usepackage{color}
\usepackage{slashed}
\usepackage{nicefrac}
\usepackage[colorlinks=true,linktocpage=true,linkcolor=blue,citecolor=blue]{hyperref}
\usepackage[utf8]{inputenc}

\long\def\comment#1{ }

\def\0{{\boldsymbol 0}}

\def\ll{{\boldsymbol l}}

\def\x{{\boldsymbol x}}

\newcommand{\beq}{\begin{eqnarray}}
\newcommand{\eeq}{\end{eqnarray}}
\newcommand{\bal}{\begin{align}}
\newcommand{\eal}{\end{align}}

\newcommand{\dd}{{\rm d}}

\begin{document}
\title{Effects of in-medium $k_T$ broadening on di-jet observables%
\thanks{Presented at XXVI Cracow EPIPHANY Conference, LHC Physics: Standard Model and Beyond}%
}
\author{Martin Rohrmoser
\address{H. Niewodnicza\'nski
Institute of Nuclear Physics
Polish Academy of Sciences, Cracow, Poland}
}
\maketitle
\begin{abstract}
We study the influence of in-medium transverse momentum broadening on observables of jet pairs produced in nuclear collisions.
These dijets were obtained numerically via the Monte-Carlo simulations of nuclear collisions, where the partons within the colliding nuclei were described via unintegrated parton densities, and their subsequent propagation within the medium of a quark-gluon plasma (QGP). 
For jet-evolution in the QGP both, kicks transverse to the momentum of the incoming particle as well as medium induced radiation were considered.   
After favorable comparison of our results with experimental LHC-data on jet-quenching we make predictions for the decorrelation of dijets. In particular, we study deviations from a transverse momentum broadening that follows a Gaussian distribution. 
\end{abstract}
  
\section{Introduction}
Jets of strongly interacting particles are abundantly produced in ultrarelativistic nuclear collisions. 
They can be loosely defined as higly energetic, collimated sprays of particles.
Due to their high energies, jets are mostly created in the initial stages of the collisions and travel through a tentative medium such as a QGP.
In particular, quark and gluon jets interact with the strongly interacting medium particles of a QGP. 
Thus, they serve as interesting probes of this kind of medium.

Experimentally, jets are studied via observables on individual jets, such as the nuclear modification factor $R_{AA}$ of jets, as well as via observables on jet pairs.
These di-jet observables also allow to study medium effects via the deviations from the momentum balance in the hard collisions. 
Specifically, we studied for this work the effects of transverse momentum transfer in the medium on angular decorrelations.
\section{Theoretical Framework}
\begin{figure}[htb]
\centerline{%
\includegraphics[width=12.5cm]{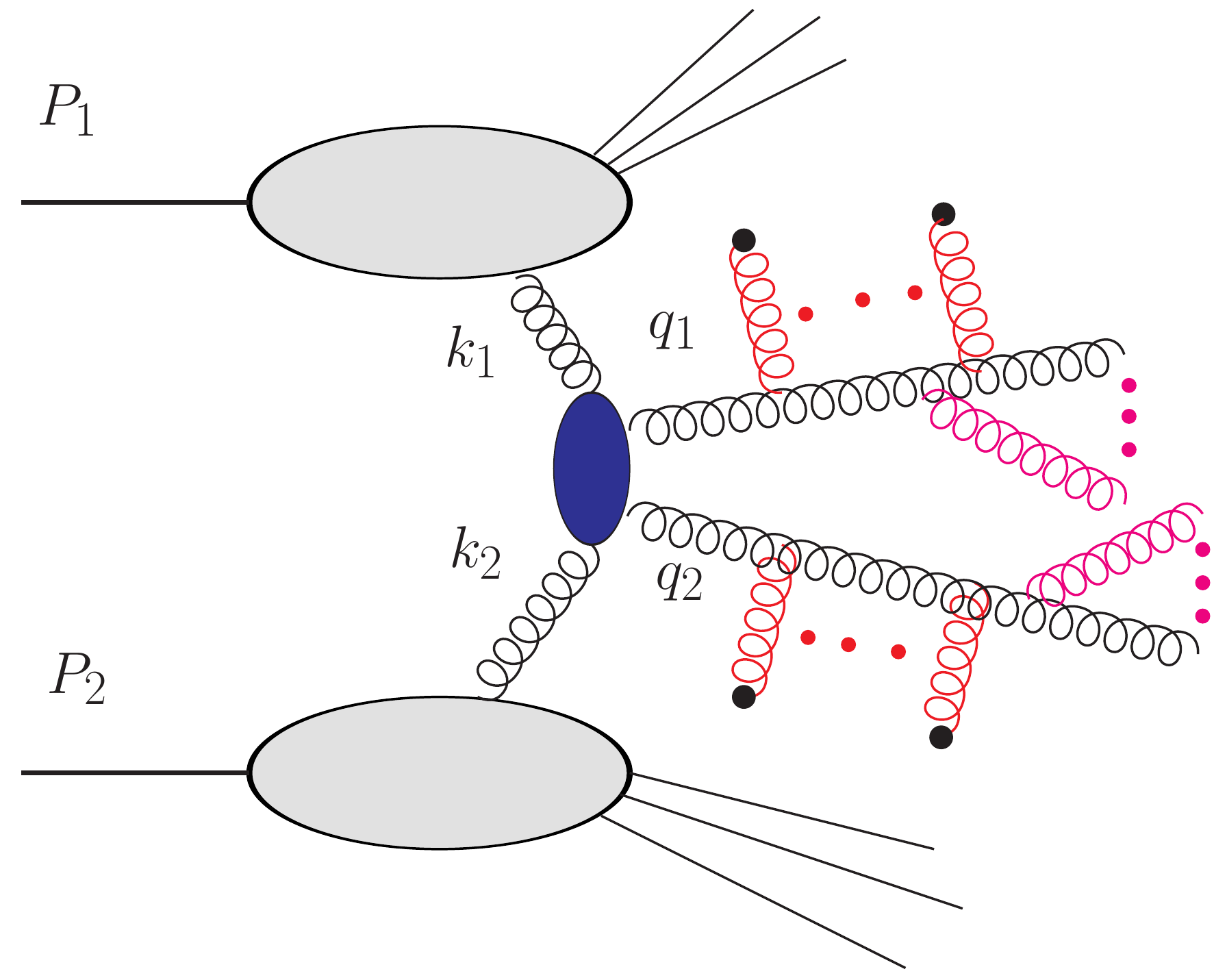}}
\caption{Schematical depiction of a nuclear collision producing two gluon jets: Nuclei with momenta $P_1$ and $P_2$ (horizontal ellipses), contain gluons with momenta $k_1$ and $k_2$ that interact in a hard collision (vertical ellipse) which produce gluons with momenta $q_1$ and $q_2$, which interact with the medium (black dots), while fragmenting into jets.}
\label{Fig:diagram}
\end{figure}

We describe the production of jet pairs in hard-nuclear collisions and the subsequent propagation of these jets through the hot and dense medium of a QGP. 
The whole process is depicted schematically in Fig.~\ref{Fig:diagram}.

In order to describe di-jet observables that depend on momentum components transverse to the beam axis, it is necessary to consider that also the partons inside the colliding nucleons have non-vanishing momentum components transverse to the beam axis,
\begin{align}
k_1=x_1\,P_1+k_{1T},&& k_2=x_2\,P_2+k_{2T}\,. 
\end{align}
where $k_1$ and $k_2$ the momenta of the partons inside the colliding nucleons with momenta $P_1$ and $P_2$, and $k_{1T}$ and $k_{2T}$ are the transverse momenta in the laboratory frame and $x_1$ and $x_2$ the momentum fractions. 
Thus, we use unintegrated parton densities rather than parton distribution functions (PDF) in order to describe the partons within the colliding nucleons. 
Therefore, we arrive at the following formula of factorization between the unintegrated parton densities and the hard-scattering,
\begin{eqnarray}\label{LO_kt-factorization} 
\frac{d \sigma_{\rm hard}}{d y_1 d y_2 d^2q_{1T} d^2q_{2T}} &=&
\int \frac{d^2 k_{1T}}{\pi} \frac{d^2 k_{2T}}{\pi}
\frac{1}{16 \pi^2 (x_1 x_2 s)^2} \; \overline{ | {\cal M}^{\mathrm{off-shell}}_{g^* g^* \to g g} |^2}
 \\  
 &&\times
{\cal F}_g(x_1,k_{1T}^2,\mu_{F}^2) \; {\cal F}_g(x_2,k_{2T}^2,\mu_{F}^2) \nonumber\\
&& \times  \delta^{2} \left( \vec{k}_{1T} + \vec{k}_{2T} 
                 - \vec{q}_{1T} - \vec{q}_{2T} \right)
\; \nonumber ,   
\end{eqnarray}
where $\sigma_{\rm hard}$ is the cross-section for the hard nuclear collision ${\cal M}^{\mathrm{off-shell}}_{g^* g^* \to g g}$ is the off-shell matrix element for the hard subprocess and ${\cal F}_g(x_i,k_{iT}^2,\mu_{F}^2)$ (with $i=1,\,2$ and $\mu_{F}$ the factorization scale) the unintegrated parton densities of gluons.
We simulated the produced jet-pairs by Monte-Carlo selection from Eq.~(\ref{LO_kt-factorization}), which was implemented in the \katie ~ program~\cite{vanHameren:2016kkz}.

In the study of the medium modifications of jets we focused on highly energetic jets, where the momentum 
of a jet particle is much larger than any transverse momentum transfer. 
Thus, one can approximate that all the transverse momentum transfers in the medium are parallel to one another.
Then, jet-medium interactions are most easily described in the reference frame where the momentum $p$ of a jet-particle (with energy E) entering the medium is given by the following light cone coordinates\footnote{This reference frame is related to the laboratory frame via a boostless rotation. For detailed description of the transformation used, cf. appendix A of~\cite{vanHameren:2019xtr}.}
\begin{align}
p_+=\sqrt{2}E\,,&& p_-=0\,,&&p_T=0\,.
\end{align}

Therefore, the medium effect on individual jet particles can be described with a fragmentation function $D$ defined as
\begin{equation}
D(\tilde{x},\ll,\tau) \equiv \tilde{x} \frac{\dd N}{\dd \tilde{x}\, \dd^2\ll} \,,
\end{equation}
where $\tilde{x}:=q_+/p_+$ (with $q_+$ defined as the light cone energy after jet-medium interaction) is the momentum fraction of in-medium energy loss and $\ll$ is the transverse momentum transfered from the medium to the jet.

For the evolution of the fragmentation function in the medium, we used an equation found by Blaizot, Dominguez, Iancu, and Mehtar Tani (BDIM)~\cite{Blaizot:2013vha},
\begin{equation}
\begin{aligned}
\frac{\partial}{\partial t} D(\tilde{x},\mathbf{l},t) = & \: \frac{1}{t^*} \int_0^1 dz\, {\cal K}(z) \left[\frac{1}{z^2}\sqrt{\frac{z}{\tilde{x}}}\, D\left(\frac{\tilde{x}}{z},\frac{\mathbf{l}}{z},t\right)\theta(z-\tilde{x}) 
- \frac{z}{\sqrt{\tilde{x}}}\, D(\tilde{x},\mathbf{l},t) \right] \\
+& \int \frac{d^2\mathbf{q}}{(2\pi)^2} \,C(\mathbf{q})\, D(\tilde{x},\mathbf{l}-\mathbf{q},t),
\end{aligned}
\label{eq:ktee1}
\end{equation}
which contains both, a medium induced splitting kernel ${\cal K}$ and a scattering kernel $C$, which describes the transverse momentum transfered by the medium. The collision kernel $C$ is 
\begin{equation}
C(\mathbf{q}) = w(\mathbf{q}) - \delta(\mathbf{q}) \int d^2\mathbf{q'}\,w(\mathbf{q'})\,,\qquad 
\label{eq:Cq}
\end{equation}
where for $w(\mathbf{q})$ we use the following definition from~\cite{Aurenche:2002pd}
\begin{equation}
 w(\mathbf{q}) = \frac{16\pi^2\alpha_s^2N_cn}{\mathbf{q}^2(\mathbf{q}^2+m_D^2)}\,,
\label{eq:wq2}
\end{equation}
where $m_D$ is the Debye mass of the medium quasi-particles.
The splitting Kernel $\cal K$ is taken as
\begin{equation}
{\cal K}(z) = \frac{\left[f(z)\right]^{5/2}}{\left[z(1-z)\right]^{3/2}}, 
\quad   f(z) = 1 - z + z^2, 
\qquad  0 \leq z \leq 1\,. 
\label{eq:kernel1}
\end{equation}
Thus, the splitting was approximated as collinear.
Furthermore, the stopping time $t^*$, which is the time after which the energy $E$ of an incoming parton is radiated off in the form of gluons, is defined as
\begin{equation}
 \frac{1}{t^*}   = \bar{\alpha}\sqrt{\frac{\hat{q}}{E}}, 
\qquad \bar{\alpha} = \frac{\alpha_s N_c}{\pi},
\label{eq:tstar}
\end{equation}
where $\hat{q}$ is the quenching parameter of the medium, $N_c$ is the number of colors and $\alpha_s$ is the QCD coupling constant.

With the above splitting kernel $\cal K$ and the stopping time $t^*$, Eq.~(\ref{eq:ktee1}) describes the medium induced radiations as coherent emissions that take into account multiple scatterings with medium particles which may occur simultaneous to the emission, and their resulting interference effects (cf.~\cite{Blaizot:2012fh}).
Thus, Eq.~(\ref{eq:ktee1}) shows the same kind of suppression for the emission of highly energetic gluons due to interferences as the approach by Baier Dokshitzer Mueller Peign\' e, Schiff, and Zakharov (BDMPS-Z)~\cite{Baier:1996kr,Baier:1996sk,Zakharov:1996fv,Zakharov:1997uu} .
In our work, jet-evolution following Eq.~(\ref{eq:ktee1}) was implemented in the Monte-Carlo program \mincas~\cite{Kutak:2018dim} . 

If one is just interested in the in-medium energy loss due to medium induced radiation, one can integrate Eq.~(\ref{eq:ktee1}) over the momentum transverse to the jet axis and obtain~\cite{Blaizot:2014rla}:
\begin{equation}
\begin{aligned}
\frac{\partial}{\partial t} D(\tilde{x},t) = & \: \frac{1}{t^*} \int_0^1 dz\, {\cal K}(z) \left[\sqrt{\frac{z}{\tilde{x}}}\, D\left(\frac{\tilde{x}}{z},t\right)\theta(z-\tilde{x}) 
- \frac{z}{\sqrt{\tilde{x}}}\, D(\tilde{x},t) \right] \,,
\end{aligned}
\label{eq:qee1}
\end{equation}
where $D(\tilde{x},t)\equiv \int \dd^2\ll\, D(\tilde{x},\ll,t)$. 
Jet-evolution in the medium following the above equation can be also calculated numerically via \mincas.

So far, the BDIM equations, Eq.~(\ref{eq:ktee1})  and Eq.~(\ref{eq:qee1}), only exists for gluons, which is why we only consider the production of pairs of gluon jets in our work. 
The resulting phenomenological predictions can, thus, only describe the qualitative behavior of di-jet observables, as we lack information on quark jets.  

We simulated a hypothetical medium by parametrizing temperature as a function of time. 
For simplicity we use the temperature dependence that follows from the Bjorken model~\cite{Bjorken:1982qr}. There, the medium temperature depends by a power law on the proper time $t$ since the creation of the medium at time $t_0$:

\begin{equation}
    T(t)=T_0\left(\frac{t_0}{t}\right)^{\frac{1}{3}}\,\qquad \textrm{with }T_0 \equiv T(t_0)\,.
\end{equation}
The temperature dependencies of the medium-properties that are necessary to describe the jet evolution following Eqs.~(\ref{eq:ktee1}) and~(\ref{eq:qee1}) can be obtained by phenomenological considerations.
Works of the JET Collaboration~\cite{Burke:2013yra} obtained the temperature dependence of the transport parameter $\hat{q}$ as 
\begin{equation}
    \hat{q}(T)=c_q T^3\,,
\end{equation}
with a yet unspecified proportionality constant $c_q$.

The number of scattering centres can be estimated by assuming a medium consisting of fermions and bosons at the thermal equilibrium, i.e. by assuming the Fermi--Dirac/Bose--Einstein distributions for the densities of quarks, anti-quarks and gluons, $n_q$, $n_{\bar{q}}$, and $n_g$, respectively. As can be shown, cf.\ e.g.\ Eq.~(3.14) in~\cite{Zapp:2008zz}, the Taylor-expansion in $T$ yields the number densities as the cubic power of $T$ at the lowest orders in $T$, so that one can write
\begin{equation}
    n(T)=n_q+n_{\bar{q}} +n_g =c_n T^3\,,
\end{equation}
with a yet unspecified proportionality constant $c_n$.

For the Debye-mass $m_D$ we assume that $m_D\propto gT$ which is consistent with findings of the Hard-Thermal-Loop (HTL) approach. In particular, following~\cite{Laine:2016hma}, we use the relation
\begin{equation}
    m_D^2=\left(\frac{N_C}{3}+\frac{N_F}{6}\right) g^2T^2\,.
\end{equation}

With respect to Ref.~\cite{Kutak:2018dim}, \mincas\ itself has not been modified, therefore the program uses constant values for the transport parameter $\hat{q}$, the number densities $n$, and the Debye mass $m_D$.
To this end, we use their average values, assuming that the jets evolve within the medium from the time-scale $t_0$ to the timescale $t_L$, i.e.

\begin{align}
    \langle \hat{q}\rangle
    &=\frac{c_q}{t_L-t_0}\int_{t_0}^{t_L} d\tau\; T_0\, \frac{t_0}{\tau}=\frac{c_q T_0^3 \ln(t_L/t_0)}{t_L/t_0-1}\,,\\
    \langle n\rangle
    &= \frac{c_n T_0^3 \ln(t_L/t_0)}{t_L/t_0-1}\,,\\
    \langle m_D\rangle
    &= \frac{3}{2}\,  \sqrt{\left(\frac{N_C}{3}+\frac{N_F}{6}\right)\frac{4\pi^2}{N_C}\bar{\alpha}}\;\;
    \frac{(t_L/t_0)^{2/3}-1}{t_L/t_0 - 1}\;T_0\,.
\end{align}

\section{Results}
\begin{figure}[htb]
\centerline{%
\includegraphics[width=12.5cm]{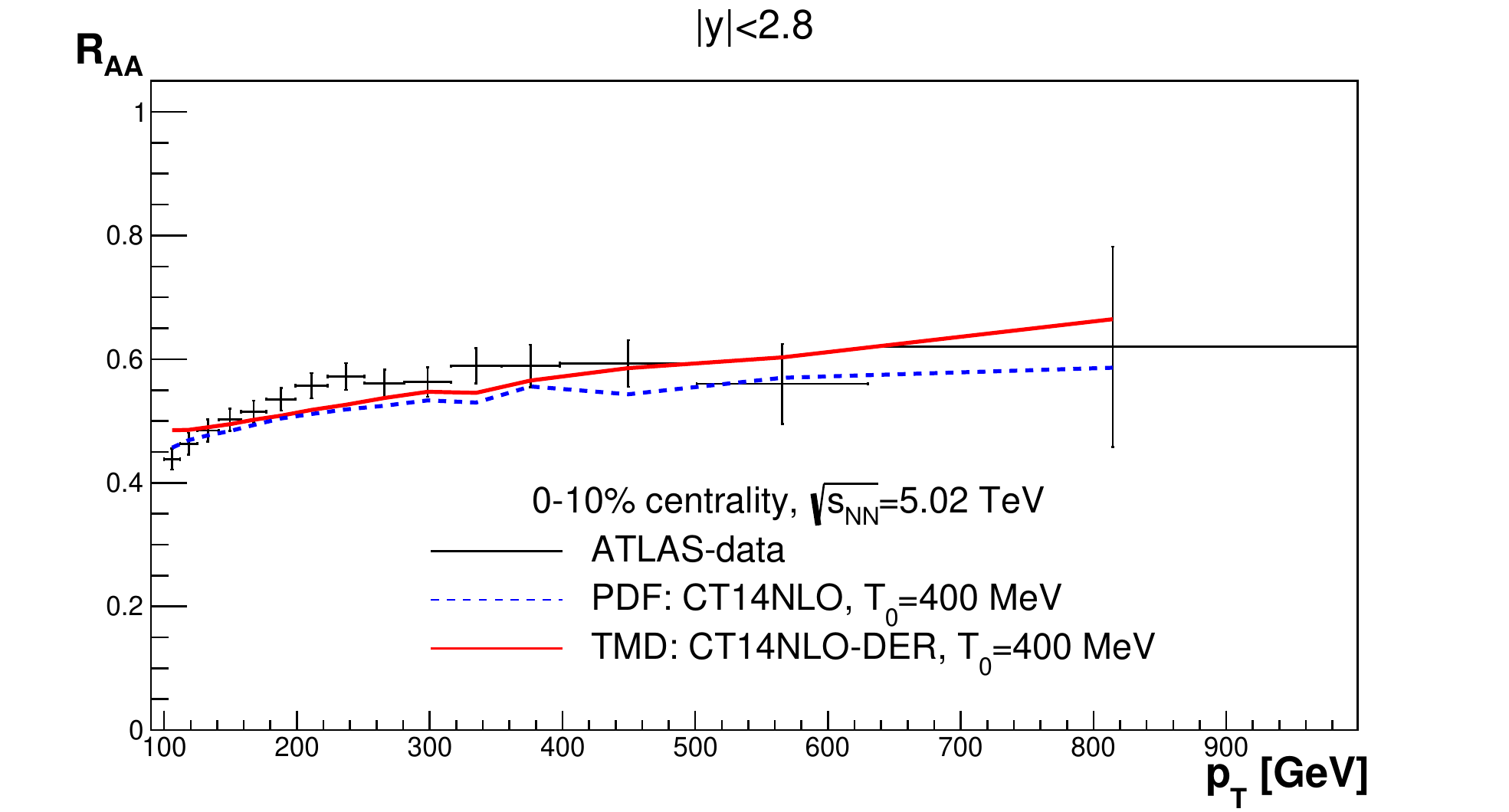}}
\caption{Nuclear modification factor $R_{\rm AA}$ at $\sqrt{s_{\rm NN}}=5.02$~TeV as predicted by \katie ~and \mincas (lines) in comparison to ATLAS data~\cite{Aaboud:2018twu} for the $10\%$ most central collisions.}
\label{Fig:RAA}
\end{figure}

In order to simulate the production of di-jets in a medium that obeys the experimentally observed jet-quenching reasonably well, we calibrated the central temperature of the medium $T_0$ to reproduce the experimentally observed nuclear modification factor $R_{\rm AA}$.
\begin{equation}
    R_{\rm AA}(p_T)=\frac{1}{\langle T_{\rm AA}\rangle}\,\frac{dN_{\rm AA}/dp_T}{d\sigma_{\rm pp}/dp_T}\,,
\end{equation}
where $\langle T_{\rm AA}\rangle$ is the average nuclear overlap function, and $\sigma_{\rm AA}$ is the cross-section for heavy ion collisions, while $\sigma_{\rm pp}$ is that of proton-proton collisions.
For the qualitative considerations of this work, nuclear effects of suppression, or enhancement, other than jet-quenching in the medium have been neglected and the nuclear modification factor is thus approximated as 
\begin{equation}
R_{\rm AA}(p_T) \approx \frac{d\sigma_{\rm AA}/dp_T}{d\sigma_{\rm pp}/dp_T}\,. 
\end{equation}
For this work, we approximated $\sigma_{\rm pp}=\sigma_{\rm hard}$ via Eq.~(\ref{LO_kt-factorization}) and obtained our numerical results by the corresponding Monte-Carlo calculations via \katie. 
The cross-section $\sigma_{\rm AA}$ for heavy ion collisions was obtained numerically by a Monte-Carlo program, which combines a selection of the jets by \katie ~and their subsequent in-medium propagation calculated via \mincas , as it was outlined in the previous section.
The results of the comparison of $R_{\rm AA}$ for jets produced in collisions at $\sqrt{s_{\rm NN}}=5.02$~TeV by our approach with the corresponding ATLAS data for the $10$\% most central collisions are shown in Fig.~\ref{Fig:RAA}.

We obtained our results using for the initial gluon distributions either the CT14 parton distribution functions (PDF) at next to leading order (NLO) in the QCD coupling constant, or the corresponding unintegrated parton densities.
As it can be seen, we were able to reproduce the experimentally observed jet-quenching reasonably well with a central temperature of $T_0=400$~MeV.
The complete set of the medium parameters is shown in Tab.~\ref{tab:parameters}.
\begin{table}[!htbp]
    \centering
    \begin{tabular}{||cc|cc|cc||}
    \hline\hline
    \multicolumn{2}{||c|}{fixed}&\multicolumn{2}{c|}{free}&\multicolumn{2}{c||}{resulting}\\\hline\hline
        $c_q$ & $3.7$ &$t_0$&$0.6\,~$fm/c & $\langle \hat{q}\rangle$& $0.54\,~$GeV$^2$/fm\\\hline
         $c_n$& $5.228$&$t_L$&$5\,~$fm/c&$\langle n\rangle$& $0.154\,~$GeV$^3$\\\hline
         && $T_0$&$0.4\,~$GeV &$\langle m_D\rangle$& 0.684~GeV\\
    \hline\hline
    \end{tabular}
    \caption{Parameters for the medium model: the parameters from theoretical/phenomenological considerations (left), the freely adjustable parameters (middle) and the resulting medium parameters used for \mincas\ (right).}
    \label{tab:parameters}
\end{table}

With the calibrated model it is then possible to study qualitatively the behavior of di-jet observables. 
In order to investigate medium effects, we simulated three different cases of di-jets
\begin{enumerate}
\item Di-jets produced in hard nuclear collisions obtained via \katie, where the individual jets undergo an in-medium propagation that follows Eq. ~(\ref{eq:ktee1}) obtained via \mincas. This case is referred to as the "Non-Gaussian $k_T$ broadening" case.
\item Di-jets produced in hard nuclear collisions obtained via \katie ~ that do not undergo any jet-medium interactions. This case is referred to as the "vacuum" case
\item Di-jets produced in hard nuclear collisions obtained via \katie , where the individual jets undergo an energy loss in the medium that follows Eq.~(\ref{eq:qee1}) obtained via \mincas. 
The medium also transfers a transverse momentum $||\ll_i||$ to the individual jets, which is selected from a Gaussian distribution, i.e.:
 \begin{equation}
        P(||\ll_i||)=\frac{1}{\sqrt{2\pi\hat{q}t_L}}\;
        \exp\left(-\frac{\ll_i^2}{2\hat{q}t_L}\right).
    \end{equation}
This case is referred to as the "Gaussian $k_T$ broadening" case.
\end{enumerate}

\begin{figure}[htb]
\centering
\includegraphics[width=12.5cm]{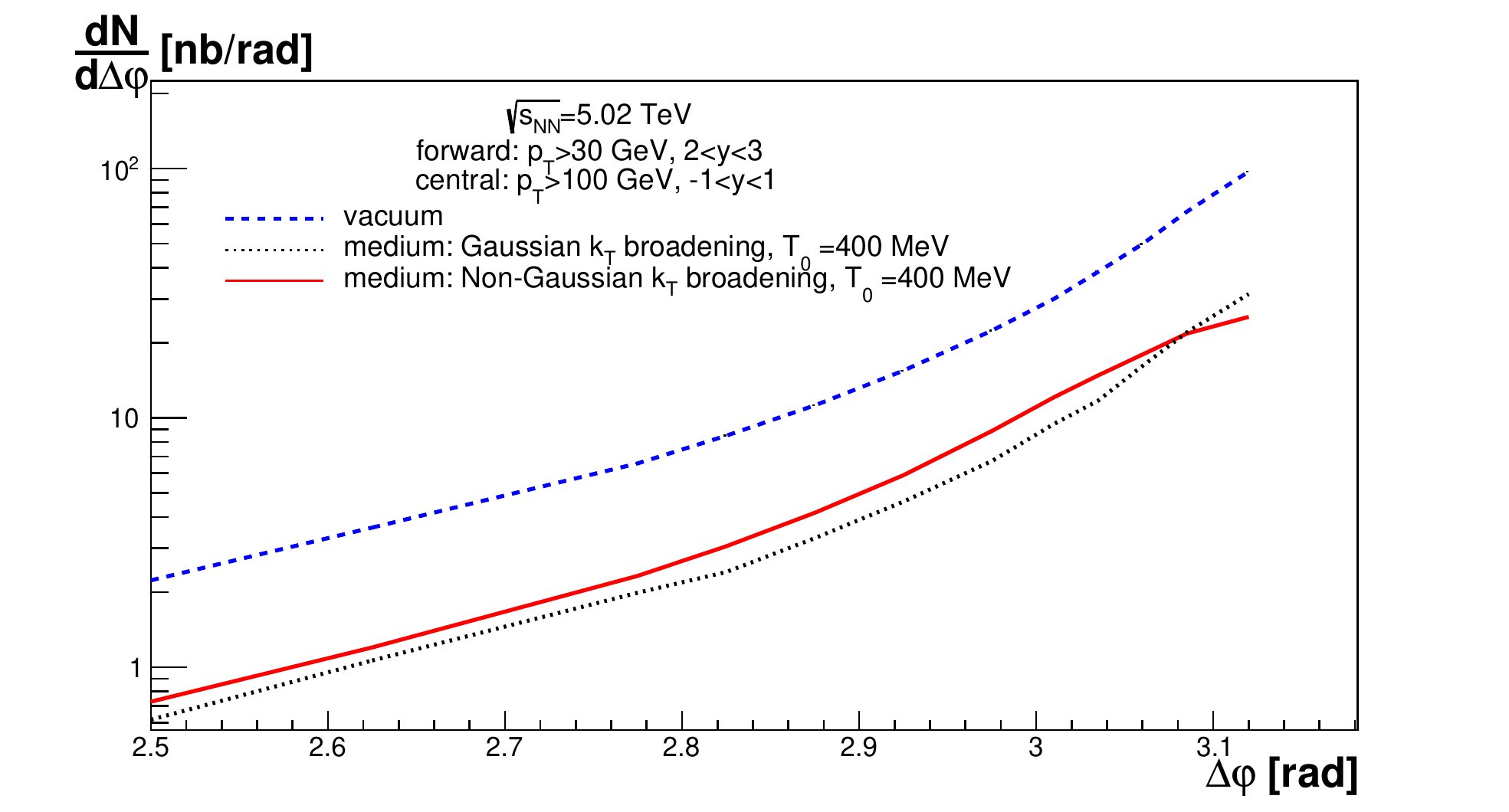}\\
\includegraphics[width=12.5cm]{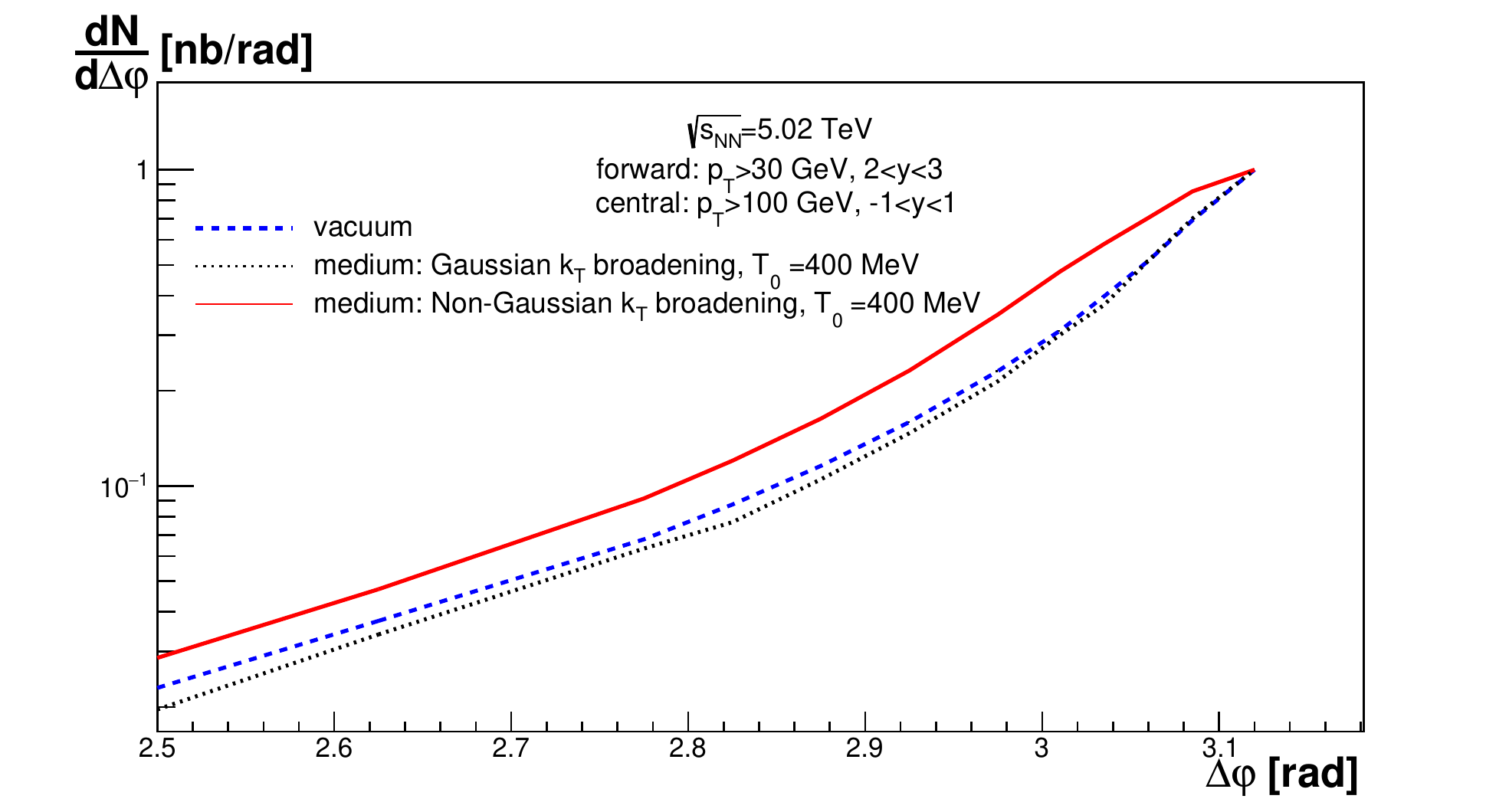}
\caption{Upper plot: Azimuthal angular decorrelations between two jets of forward and central rapidities without medium modifications (dashed line), with Gaussian $k_T$ broadening (dotted line) and Non-Gaussian $k_T$ broadening (solid line). Lower plot: Same as the upper one, but normalized to the maximum of the distribution}
\label{Fig:dphi}
\end{figure}

In particular, we focused on the behavior of the azimuthal decorrelations of di-jets, defined as $dN/d\Delta \phi$, the number of jet-pairs $N$, where the momenta of the two jets differ in their azimuthal angles by the value $\Delta \phi$.
Numerical results for azimuthal decorrelations of di-jets produced in nuclear collisions at $\sqrt{s_{NN}}=5.02$~TeV with one jet going into a forward and the other one in a central rapidity direction ($2<y<3$ with $p_t>30$~GeV and $-1<y<1$ with $p_t>100$~GeV, respectively) are shown in the upper panel of Fig.~\ref{Fig:dphi}. 
As it can be seen the medium effects lead to a considerable suppression of the observed jet-pairs, by at least a factor of $3$ in the region $2.5<\Delta\phi<\pi$ for both results from both cases, Gaussian and the Non-Gausian $k_T$ broadening.
Furthermore, the distribution for Non-Gaussian $k_T$ broadening appears to be slightly broader than the distribution for Gaussian $k_T$ broadening.
The different widths of the distributions can be seen more easily, when the distributions are normalized to their respective maxima, as it is shown in the lower panel of Fig.~\ref{Fig:dphi}.
As it can be seen, the distributions for Gaussian $k_T$ broadening and the vacuum case show a very similar behavior, while the distribution for the Non-Gaussian $k_T$ broadening is considerably broader.

\section{Summary and conclusions}
We have simulated the production of di-jets in heavy ion collision in a Monte-Carlo approach.
This approach simulates di-jet production from hard nuclear collisions, where the incoming partons are described via unintegrated parton densities, with the KATIE program, and the subsequent propagation of these jets in the medium, subjected to both, in medium transverse momentum broadening as well as medium induced radiation, which was simulated by the MINCAS program, which describes jet-evolution following the BDIM-equation.
Due to the use of unintegrated parton densities for the incoming partons, the observables of azimuthal angular decorrelation of di-jets could be studied and possible medium effects on these observables were investigated. 

We found that the azimuthal decorrelations show a considerable broadening due to medium effects. 
In particular, we studied the dependence of this effect on the transverse momentum $k_T$ that is transferred from the medium to an individual jet. 
We found that the deviation of the $k_T$ distributions from a Gaussian distribution is mainly responsible for the broadening of the azimuthal di-jet decorrelations.

So far, our approach only considers pairs of gluon jets, which are the dominant contributions to jets in the mid-forward rapidity region, which was investigated in this work. 
In the future, we would like to extend our approach to even more forward rapidity regions on the one hand, and to the inclusion of quark-jets on the other hand.
\section*{Acknowledgements}
These conference proceedings are based on common work~\cite{vanHameren:2019xtr} with Andreas van Hameren, Krzysztof Kutak, Wies\l{}aw P\l{}aczek, and Konrad Tywoniuk.
I thank them for their contributions.
The research is supported by grant  
Polish National Science Centre grant no. DEC-2017/27/B/ST2/01985
\providecommand{\href}[2]{#2}
\end{document}